\pdfoutput=1
\documentclass[11pt,a4paper]{article}
\usepackage{amsmath,amssymb,booktabs,enumerate}
\usepackage{hyperref}
\usepackage[dvipdfmx]{graphicx}
\usepackage{url,cite}
\usepackage[utf8]{inputenc}
\usepackage{comment}
\usepackage{color}

\def\beq{\begin{eqnarray}}
\def\eeq{\end{eqnarray}}
\def\be{\begin{eqnarray}}
\def\ee{\end{eqnarray}}
\def\no{\nonumber}

\begin{document}

\begin{center}
{\large \bf O($N$) Invariance of the Multi-Field Bounce
}
\vskip 1cm
Kfir Blum$^1$$^\dag$,
Masazumi Honda$^1$,
Ryosuke Sato$^1$,\\
Masahiro Takimoto$^{1,2}$,
and Kohsaku Tobioka$^{1,3}$
\vskip 0.4cm

{\it
$^1$ Department of Particle Physics and Astrophysics,\\
Weizmann Institute of Science, Rehovot 7610001, Israel\\
$^2$ Theory Center, 
High Energy Accelerator Research Organization (KEK), \\
Tsukuba 305-0801, Japan\\
$^3$ Raymond and Beverly Sackler School of Physics and Astronomy,\\
Tel-Aviv University, Tel-Aviv 6997801, Israel
}
\vskip 1.5cm

\abstract{
In his 1977 paper on vacuum decay in field theory: {\it The Fate of the False Vacuum}, Coleman considered the problem of a single scalar field and assumed that the minimum action tunnelling field configuration, the {\it bounce}, is invariant under O(4) rotations in Euclidean space. A proof of the O(4) invariance of the bounce was provided later  by Coleman, Glaser, and Martin (CGM), who extended the proof to $N>2$ Euclidean dimensions but, again, restricted non-trivially to a single scalar field. As far as we know a proof of O($N$) invariance of the bounce for the tunnelling problem with multiple scalar fields has not been reported in the QFT literature, even though it was assumed in many works since. We make progress towards closing this gap. 
Following CGM we define the reduced problem of finding a field configuration minimizing the kinetic energy at fixed potential energy. Given a solution of the reduced problem, the minimum action bounce can always be obtained from it by means of a scale transformation. We show that if a solution of the reduced problem exists, then it and the minimum action bounce derived from it are indeed O($N$) symmetric. We review complementary results in the mathematical literature that established the existence of a minimizer under specified criteria.  
}
\end{center}

\section{Introduction and result}\label{sec:int}
In his 1977 paper on vacuum instability in field theory, Coleman~\cite{Coleman:1977py}  considered the problem of a single scalar field and assumed that the tunnelling field configuration of minimum Euclidean action, the {\it bounce}, is O(4)-invariant. A proof of the O(4) invariance of the bounce was provided later in Ref.~\cite{Coleman:1977th}. This proof was given for $N>2$ Euclidean dimensions, and was restricted non-trivially to the case of a single scalar field. 

As far as we know, a proof of O($N$) invariance of the bounce with multiple scalar fields has not been reported in the QFT literature, although it was assumed implicitly or explicitly in many works (for a handful of examples, see Refs.~\cite{Sher:1988mj,Kusenko:1996jn,Dienes:2008qi,Aravind:2014pva,Aravind:2014aza,Greene:2013ida,Dine:2015ioa,Andreassen:2016cvx,Kusenko:1995bw,Kusenko:1996bv}). 
The purpose of the current paper is to make progress towards closing this gap.

Following Ref.~\cite{Coleman:1977th} we define the reduced problem of finding a field configuration minimizing the kinetic energy at fixed  potential energy. Given a solution of the reduced problem, it is known that the minimum action bounce can always be obtained from it by means of a scale transformation.
We show that {\it if a solution of the reduced problem exists, then it and the minimum action bounce derived from it are O($N$) symmetric}. 

The paper is organised as follows. Sec.~\ref{ssec:A} reviews the definition of the reduced problem introduced in Ref.~\cite{Coleman:1977th} and {\it theorem A} that was proved in that paper and that is used in our analysis. Sec.~\ref{ssec:O(N)} is our main contribution, showing that the solution of the reduced problem, if it exists, possesses O($N$) symmetry. As in Ref.~\cite{Coleman:1977th} we restrict to $N>2$ dimensions. 
Sec.~\ref{sec:math} reviews complementary results in the math literature, some of which established the existence of a minimizer under suitable admissibility criteria, that we relate to phenomenologically interesting QFTs. In Sec.~\ref{sec:conc} we conclude.

\section{The reduced problem}\label{ssec:A}
First we recall some preliminaries.
We are interested in the scalar multi-field configuration $\Phi=\left\{\Phi_a\right\}$, $a=1,...,m$. The Euclidean equations of motion (EOM) are
\be\label{eq:eom}\sum_{i=0}^{N-1}\frac{\partial^2\Phi_a}{\partial x_i^2}-\frac{\partial U}{\partial\Phi_a}=0,\ee
where $U(\Phi)$ is the potential energy density.  
%
We make the following admissibility assumptions about $U$:
\begin{description}
\item[(A1)] $U$ is continuously differentiable everywhere in field space,
\item[(A2)] $U(0)=\partial U/\partial\Phi_a|_{\Phi=0} = 0$,  
\item[(A3)] $U$ is somewhere negative,
\item[(A4)] $U$ is stabilized at the origin. For concreteness we impose that all of the eigenvalues of the Hessian of $U$ at $\Phi=0$ are positive.
\end{description}

The kinetic and potential energy functionals associated with $\Phi$ are given by
\be T[\Phi]&=&\int d^Nx\sum_{a=1}^m\sum_{i=0}^{N-1}\frac{1}{2}\left(\frac{\partial\Phi_a}{\partial x_i}\right)^2,\\
V[\Phi]&=&\int d^NxU(\Phi).
\ee
The action is
\be S=T+V.\ee
%


We define a scale transformation by~\cite{Derrick:1964ww}
\be\label{eq:trans}\Phi_a^{(\sigma)}(x)=\Phi_a(x/\sigma),\ee
where $\sigma$ is a positive number. Then $V$ and $T$ transform as
\be\label{eq:st} V[\Phi^{(\sigma)}]=\sigma^NV[\Phi],\;\;\;\;T[\Phi^{(\sigma)}]=\sigma^{N-2}T[\Phi].\ee
Any solution of Eq.~(\ref{eq:eom}) makes $S$ stationary. In particular $S$ must be stationary w.r.t.~scale transformations. This leads to 
\be\label{eq:vt} (N-2)T+NV=0,\ee
or equivalently
\be\label{eq:vt2} S=\frac{2T}{N},\ee
for any solution of Eq.~(\ref{eq:eom}).
A non-trivial solution of Eq.~(\ref{eq:eom}) has $T>0$ and, by Eq.~(\ref{eq:vt}), $V<0$.\\
 
We define the reduced problem as the problem of finding a collection of configurations $\Phi_a$ vanishing at infinity\footnote{``Vanishing at infinity" means that for any positive number $\epsilon$ the set of all points for which $|\Phi_a|\geq\epsilon$ has finite Lebesgue measure.} which minimizes $T$ for some fixed negative $V$.

If a solution of the reduced problem is found for some negative $V$, then by applying the appropriate scale transformation we can find a solution for any negative $V$. To see this, consider the scale-invariant quantity
\be R&=&\frac{T^{\frac{N}{N-2}}}{-V}.\ee
For fixed negative $V$, minimizing $R$ is equivalent to minimizing $T$. 
However, all configurations that are scale-transformed of each other have the same value of $R$. Thus the reduced problem can equivalently be stated as the problem of finding a configuration with arbitrary negative $V$ that minimizes $R$.\\ 

{\it Theorem A: if a solution of the reduced problem exists, then, for appropriately chosen $V$, it is a solution of Eq.~(\ref{eq:eom}) that has action less than or equal to that of any non-trivial solution of Eq.~(\ref{eq:eom}).} 

A proof of Theorem A was given in Ref.~\cite{Coleman:1977th}. In the rest of this section we review this proof. 
The first step is to show that a solution of the reduced problem can always be scale-transformed into a solution of Eq.~(\ref{eq:eom}). A solution $\tilde\Phi$ of the reduced problem stationarizes 
\be\label{eq:Stilde} S'[\tilde\Phi]=T[\tilde\Phi]+\lambda^2(V[\tilde\Phi]-V_0),\ee
where $V_0<0$ is a negative number and $\lambda^2$ is a Lagrange multiplier.
Stationarity w.r.t.~scale transformations yields
\be\label{eq:vtp}(N-2)T[\tilde\Phi]+\lambda^2NV[\tilde\Phi]=0.\ee
Since $V[\tilde\Phi]$ is negative and $T[\tilde\Phi]$ is positive we have $\lambda^2>0$, and we can define the scale-transformed configuration $\Phi(x)=\tilde\Phi^{(\lambda)}(x)=\tilde\Phi(x/\lambda)$. The equation of motion obeyed by $\tilde\Phi(x)$ is the same as Eq.~(\ref{eq:eom}) with the replacement $\frac{\partial U}{\partial\Phi_a}\to\lambda^2\frac{\partial U}{\partial\tilde\Phi_a}$. Using this it is easy to verify that $\Phi$ satisfies Eq.~(\ref{eq:eom}).

The second step is to show that the solution $\Phi$ constructed above has $S$ less than or equal to that of any solution of Eq.~(\ref{eq:eom}). Let $\bar\Phi$ be a non-trivial solution of Eq.~(\ref{eq:eom}). 
Now, let $\tilde\Phi$ be a solution of the reduced problem with $V[\tilde\Phi]=V[\bar\Phi]$. By the definition of the reduced problem, $T[\tilde\Phi]\leq T[\bar\Phi]$ which, comparing Eqs.~(\ref{eq:vt}) and~(\ref{eq:vtp}), gives $\lambda\leq1$.
Proceeding as before, $\Phi=\tilde\Phi^{(\lambda)}$ satisfies Eq.~(\ref{eq:eom}), but with
\be\label{eq:typo} T[\Phi]=\lambda^{(N-2)} T[\tilde\Phi]\leq T[\tilde\Phi].\ee
Using Eq.~(\ref{eq:vt2}) we finally have 
\be S[\Phi]\leq S[\bar\Phi],\ee
where equality holds if and only if $\bar{\Phi}$ is a solution of the reduced problem. This completes the proof of Theorem A. 
As noted in Ref.~\cite{Coleman:1977th}, the proof holds for an arbitrary number of scalar fields $m$.

\section{O($N$) invariance}\label{ssec:O(N)}
Let us assume that there exists a multi-field configuration that solves the reduced problem. 
By Theorem A we can use a scale transformation to construct a solution $\Phi$ of Eq.~(\ref{eq:eom}), with negative $V[\Phi]$ and with action $S[\Phi]$ that is equal to or smaller than that of any non-trivial solution of Eq.~(\ref{eq:eom}). 
Furthermore, $R[\Phi]$ is equal to or smaller than $R$ of any other configuration with negative $V$ (strictly smaller if the other configuration is not a solution of the reduced problem).\\

The following chain of arguments shows that $\Phi$ possesses O($N$) symmetry.\\

We choose a Cartesian coordinate system $\{x_0,x_1,...,x_{N-1}\}$ and pay particular attention to the $x_0$ direction. The choice of coordinate system and of $x_0$ is arbitrary. 
Define the kinetic and potential   surface energy densities,
\be
T(t)&=&\int d^{N-1}x\sum_{a=1}^m\sum_{i=0}^{N-1}\frac{1}{2}\left(\frac{\partial\Phi_a}{\partial x_i}\right)^2 \biggr|_{x_0=t}\\
V(t)&=&\int d^{N-1}xU(\Phi) |_{x_0=t}\ee
where $d^{N-1}x=dx_1dx_2...dx_{N-1}$. Of course,
\be T=\int dtT(t),\;\;\;V=\int dtV(t).\ee
Let us consider a surface $x_0=t$, dividing space into two parts $x_0>t$ and $x_0<t$. For each part, the kinetic and potential energy are given by
\be T^t_+&=&\int_t^\infty dt'T(t'),\;\;\;V^t_+=\int_t^\infty dt'V(t'),\\
T^t_-&=&\int^t_{-\infty} dt'T(t'),\;\;\;V^t_-=\int^t_{-\infty} dt'V(t'),
\ee
such that 
\be
T^t_++T^t_-=T,\;\;\;\;V^t_++V^t_-=V.\ee

Now, let us construct a field configuration by reflecting the region $x_0>t$ onto the region $x_0<t$. We call this configuration $\Phi^t_+$. To be precise, we define
%
\be\Phi^t_+&=& \Phi(t+|x_0-t|,x_1,...).\ee
Analogously, we also construct the opposite reflection $\Phi^t_-$. 
The reflected configurations satisfy\footnote{Strictly speaking, $\partial\Phi^t_\pm/\partial x_0$ and thus the kinetic energy density associated with it are undefined at the point $x_0=t$. However, the discontinuity is integrable and the kinetic and potential energy are well behaved everywhere.
}:
\be T[\Phi^t_{\pm}]&=&2T^t_{\pm},\label{eq:Tt}\\
V[\Phi^t_{\pm}]&=&2V^t_{\pm},\label{eq:Vt}\\
R[\Phi^t_\pm]&=&\frac{\left(2T^t_{\pm}\right)^{\frac{N}{N-2}}}{-2V^t_{\pm}}.\label{eq:Rt}\ee

$V(t)$ is a continuous function. Thus there exists $t^*$ for which
\be\label{eq:tsV} V^{t^*}_{+}=V^{t^*}_{-}=\frac{V}{2}.\ee
Since $\Phi$ is a solution of the reduced problem, $R[\Phi]\leq R[\Phi^t_\pm]$ for any $t$.
Therefore,
\be\label{eq:tsT} T^{t^*}_{+}=T^{t^*}_{-}=\frac{T}{2}.\ee
Otherwise, either $R[\Phi^{t^*}_+]<R[\Phi]$ or $R[\Phi^{t^*}_-]<R[\Phi]$.

Let us, for the sake of clarity, redefine the $x_0$ coordinate setting $t^*=0$. 
We then construct an infinitesimal perturbation by considering the surface $x_0=\epsilon/2$ with sufficiently small $\epsilon$.
We have
\be
T[\Phi_-^{\epsilon/2}]&=&T+2\int_0^{\epsilon/2}dtT(t)=T+\epsilon T(0),\\
V[\Phi^{\epsilon/2}_-]&=&V+\epsilon V(0),\\
T[\Phi^{\epsilon/2}_+]&=&T-\epsilon T(0),\\
V[\Phi^{\epsilon/2}_+]&=&V-\epsilon V(0).
\ee
Computing $R$ for the deformed configurations, we have
\be R[\Phi^{\epsilon/2}_-]&=&R[\Phi]+\epsilon R[\Phi]\left(\frac{NT(0)}{(N-2)T}-\frac{V(0)}{V}\right)=R[\Phi]-\epsilon \frac{R[\Phi]}{V}\left(T(0)+V(0)\right),\\
R[\Phi^{\epsilon/2}_+]&=&R[\Phi]+\epsilon \frac{R[\Phi]}{V}\left(T(0)+V(0)\right),\ee
where we made use of Eq.~(\ref{eq:vt}). 
Imposing $R[\Phi^{\epsilon/2}_-]\geq R[\Phi]$ and $R[\Phi^{\epsilon/2}_+]\geq R[\Phi]$ we obtain
\be\label{eq:TV0} V(0)+T(0)=0.\ee

We gain more mileage from Eq.~(\ref{eq:TV0}) as follows. Acting with $\int d^{N-1}x\sum_a\frac{\partial\Phi_a}{\partial x_0}$ on the EOM and integrating by parts, we have
\be0&=&\int d^{N-1}x\sum_a\frac{\partial\Phi_a}{\partial x_0}\left[\frac{\partial^2\Phi_a}{\partial x_0^2}+\sum_{j=1,2,...}\frac{\partial^2\Phi_a}{\partial x_j^2}-\frac{\partial U}{\partial\Phi_a}\right]\no\\
&=&\frac{\partial}{\partial x_0}\int d^{N-1}x\left[\sum_a\left\{\left(\frac{\partial\Phi_a}{\partial x_0}\right)^2-\sum_{i=0,1,2,...}\frac{1}{2}\left(\frac{\partial\Phi_a}{\partial x_i}\right)^2\right\}-U\right].\ee
Notice that the sum on $i$ in the second line includes $i=0$. Since all fields and derivatives vanish at $|x_0|\to\infty$, the quantity on which $\partial_{x_0}$ acts is zero at any $x_0$, implying
\be\label{eq:consx0} \int d^{N-1}x\sum_a\left(\frac{\partial\Phi_a}{\partial x_0}\right)^2 \biggr|_{x_0=t}&=&T(t)+V(t)\ee
for any $t$. 
Combining Eqs.~(\ref{eq:TV0}) and~(\ref{eq:consx0}) we find that
\be\label{eq:refx0} \int d^{N-1}x\sum_a\left(\frac{\partial\Phi_a}{\partial x_0}\right)^2\biggr|_{x_0=0}&=&0.\ee
Therefore the first derivative of all of the $\Phi_a$ w.r.t.~$x_0$ vanishes on the  $N-1$ dimensional surface defined by $x_0=0$.\\ 

The surface $x_0=t^*$ (which we took to be $t^*=0$) is unique: there is no other parallel surface $x_0={\tilde t}^*$, with ${\tilde t}^*\neq t^*$, at which Eq.~(\ref{eq:refx0}) is satisfied. If there were another ${\tilde t}^*\neq t^*$, say ${\tilde t}^*<t^*$, then the contribution to the kinetic energy from the interval $({\tilde t}^*,t^*)$ must vanish, implying that $\Phi_a=0$ in the interval. In that case we could construct a new configuration $\tilde\Phi$ by clipping $\Phi$ at $x_0<t^*$, namely, $\tilde\Phi(\vec x)=\theta(x_0)\Phi(\vec x)$, where $\theta(x)$ is the Heaviside step function. A quick calculation shows that $R[\tilde\Phi]=\left(2^{-\frac{2}{N-2}}\right)R[\Phi]<R[\Phi]$, in contradiction with $\Phi$ being a solution of the reduced problem.\\

The choice of coordinate system and of $x_0$ in Eq.~(\ref{eq:refx0}) is arbitrary. Therefore for any direction $\hat n$ we have a unique surface orthogonal to $\hat n$ across which the first derivative of all of the $\Phi_a$ vanishes. We denote such surface an $\hat n^*$-surface. 

For clarity we divide the following final arguments into 4 steps. \\

{\bf Step 1.} Here we create an $N$-fold parity symmetric
solution of the reduced problem $\Phi^{P_N}$ based on the original configuration $\Phi$.
First, we choose a coordinate system $\{x_0,x_1,...,x_{N-1}\}$. 
Next, we fold $\Phi$ $N$ times by reflecting, for example, first the region $x_0>0$ onto the region $x_0<0$ after adjusting $\hat x_0^*$-surface at $x_0=0$, then the region $x_1>0$ onto $x_1<0$, and so on. 
The configuration after $N$ reflections is denoted $\Phi^{P_N}$. 
It is easy to see that $\Phi^{P_N}$ has mirror symmetries (parity) across all of the $\hat x_i^*$ surfaces 
and is a solution of the reduced problem.\\

{\bf Step 2.}
The uniqueness of $x_0=t^*=0$ isolates the point $\vec x=0$ as the intersection of the $N$ reflection surfaces of $\Phi^{P_N}$. 
The point $\vec x=0$ is the physical centre of the bounce. It is easy to see that the centre of the bounce is unique, namely that any reflection surface (orthogonal to some arbitrary direction $\hat n$) must pass through $\vec x=0$. To see this, consider a surface orthogonal to some direction $\hat n$ that is a linear combination of the original $\hat x_i$. If the new surface does not pass through $\vec x=0$, say it is displaced from the origin by an impact parameter $d$, then by a combination of $N$ reflections across the original $\hat x_i$ axes we can construct a new surface parallel to the first one and displaced from it by $2d$ along $\hat n$, in contradiction with the uniqueness of the reflection surface per direction $\hat n$.\\ 

\begin{figure}[t]
\begin{center}
\includegraphics[scale=0.13]{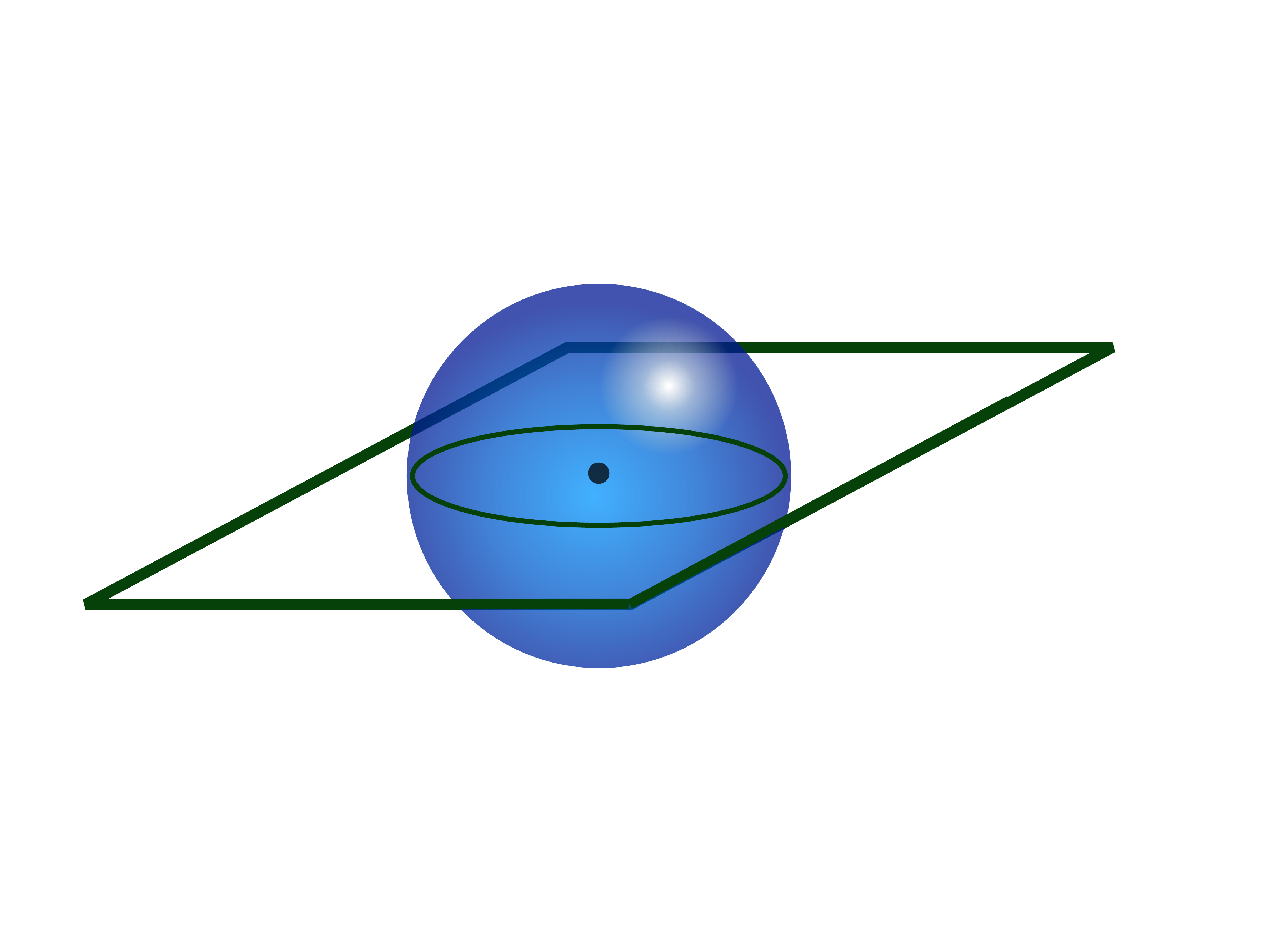}
\includegraphics[scale=0.13]{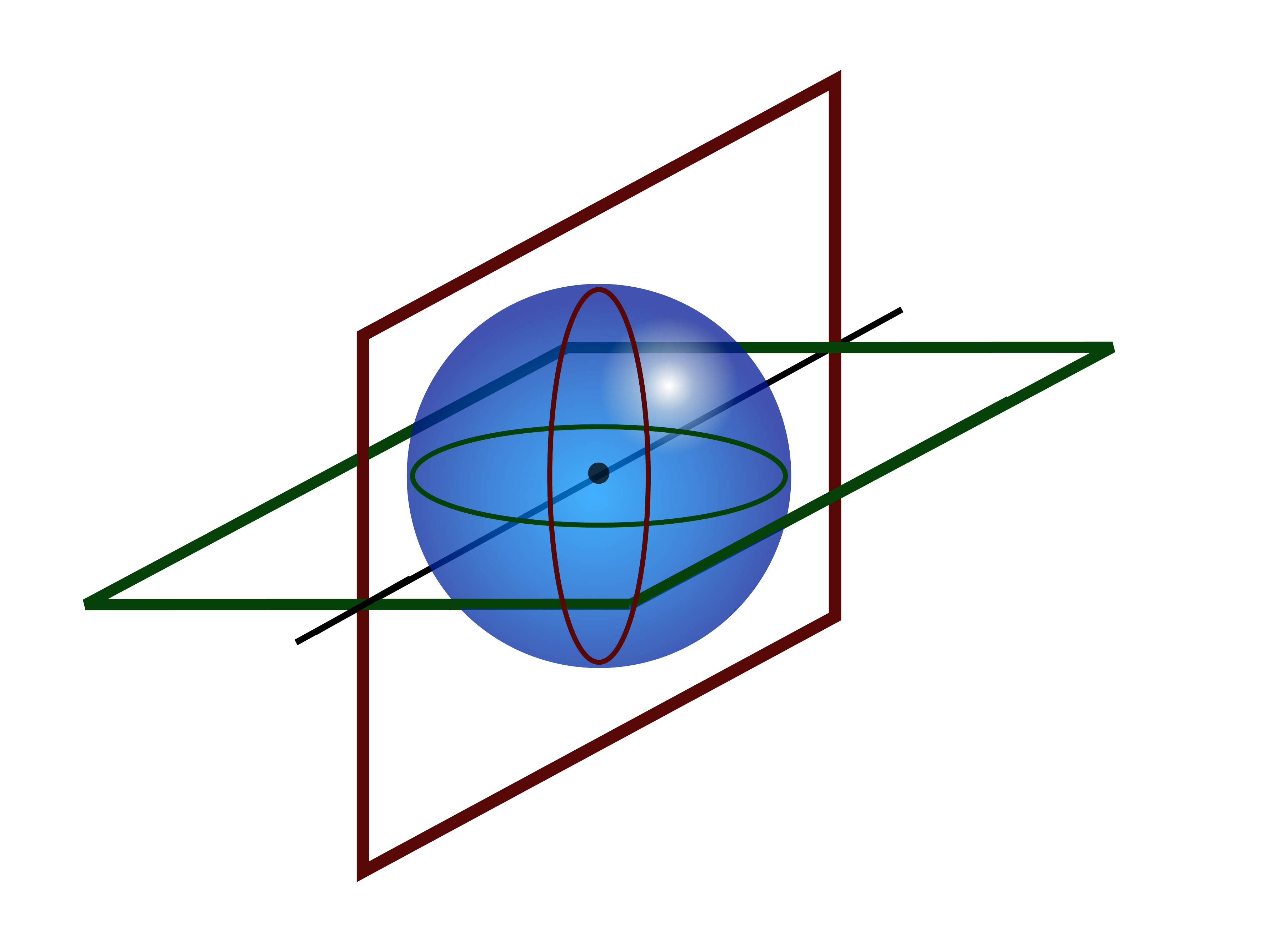}
\includegraphics[scale=0.13]{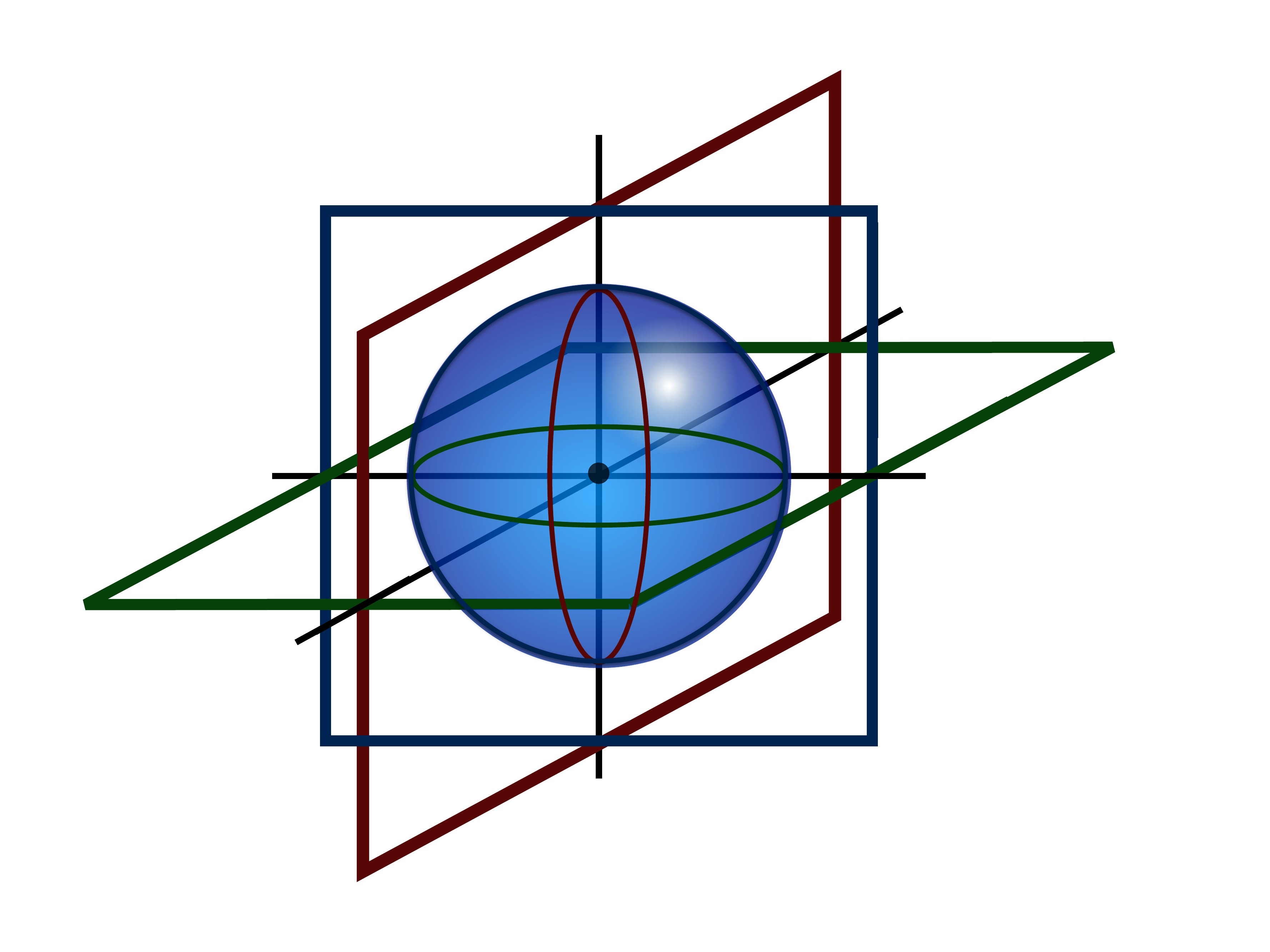}
\caption{One, two, three orthogonal $t^*$-surfaces for a bounce in N=3.}
\label{fig:cut}
\end{center}
\end{figure}

{\bf Step 3.} $\Phi^{P_N}$ is invariant to O($N$) rotations around $\vec x=0$. Consider $\Phi^{P_N}_a(\vec y)$ for some arbitrary point $\vec y$. An infinitesimal O($N$) rotation takes $\Phi^{P_N}_a(\vec y)\to\Phi^{P_N}_a(\vec y+\epsilon\hat n)$, where $\hat n\cdot\vec y=0$. Assigning the coordinate $x_0$ to the direction $\hat n$, the coordinate $x_1$ to $\hat y$, and using Eq.~(\ref{eq:refx0}) we find 
\be\delta\Phi^{P_N}_a&=&\epsilon\left(\frac{\partial\Phi^{P_N}_a}{\partial x_0}\right)_{x_0=0}=0.
\ee
Thus $\Phi^{P_N}$ is O($N$) invariant. Fig.~\ref{fig:cut} illustrates the construction for $N=3$.
\\

{\bf Step 4.}
Finally, we show the O($N$) invariance of the original configuration $\Phi$.
From step 3 we know that after $N$ reflection operations,
the original $\Phi$ becomes the O($N$) symmetric $\Phi^{P_N}$.
Take one step back and consider the configuration $\Phi^{P_{N-1}}$ obtained after $N-1$ reflections. 
Note that from $\Phi^{P_{N-1}}$ we obtain an $N$-fold parity invariant configuration, and therefore an O($N$) symmetric configuration, both if we reflect the region $x_{N-1}>0$ onto the region $x_{N-1}<0$ or vice-verse, $x_{N-1}<0$ onto $x_{N-1}>0$. Given a continuously differentiable $U$ we know that solutions of the reduced problem are continuous\footnote{A solution of the reduced problem satisfies an elliptic differential equation given by Eq.~(\ref{eq:eom}) with $\left(\partial U/\partial\Phi_a\right)\to\lambda^2\left(\partial U/\partial\Phi_a\right)$ with $\lambda^2>0$, and so it is continuous for continuously differentiable $U$.}. From continuity it follows that $\Phi^{P_{N-1}}$ must already be O($N$) symmetric. 
Tracking the argument $N-1$ times backwards we conclude that the original configuration
$\Phi$ is O($N$) symmetric.

\section{Complementary results in the math literature, and some examples}\label{sec:math}
A proof of O($N$) invariance of the solution of a functional minimization problem equivalent to our reduced problem was given by Lopes~\cite{Lopes1996378}, albeit without reference to action extremization. While it differs in details, the basic construction in~\cite{Lopes1996378} resembles ours: identifying hyper-surfaces that divide equally the potential and kinetic energy of the field configuration. More recently, Ref.~\cite{2008arXiv0806.0299B} presented a proof that parallels ours (though, again, differing in details) and extends to $N=2$, and discussed the connection to action extremization via the scaling argument. 

Our proof of $O(N)$ symmetry (and likewise the proofs in~\cite{2008arXiv0806.0299B,Lopes1996378}) assumes the existence of a solution -- a {\it minimizer} -- of the reduced problem. An important caveat is that, in some cases, a minimizer may not exist. 
Ref.~\cite{brezis1984} addressed this problem, without attending to the question of radial symmetry. For $N\geq3$, the existence of a minimizer was established for continuous $U(\Phi)$ satisfying $U(0)=0$, subject to the following additional conditions\footnote{We choose to work with (2.6-2.8) of~\cite{brezis1984}, rather than their (2.5).}:
\begin{description}
\item[(R1)] $\displaystyle{\limsup_{|\Phi|\to\infty}\,|\Phi|^{-\frac{2N}{N-2}}\,U(\Phi)\geq0}$,
\item[(R2)] $\displaystyle{\limsup_{|\Phi|\to0}\,|\Phi|^{-\frac{2N}{N-2}}\,U(\Phi)\geq0}$,  
\item[(R3)] $U$ is somewhere negative,
\item[(R4)] {\bf (i)} $\displaystyle{\limsup_{|\Phi|\to\infty}\,|\Phi|^{-\frac{2N}{N-2}}\,|U(\Phi)|=0}$, \\{\bf or}\\
{\bf (ii)} $\displaystyle{U\in C^1\left(R^m\setminus\{0\}\right)}$ {\bf and} $\displaystyle{\left|\nabla U(\Phi)\right|\leq C+C|\Phi|^{\frac{N+2}{N-2}}}$,\\
{\bf or}\\
{\bf (iii)} $\displaystyle{U\in C^1\left(R^m\setminus\{0\}\right)}$ {\bf and} 
$\displaystyle{\left|\nabla U(\Phi)\right|\leq C+C|\Phi|^{q-1}}$ {\bf and} 
$|\Phi|^{\frac{2N}{N-2}}+|U(\Phi)|\geq\alpha|\Phi|^q-C$, 
\end{description}
where $\alpha,C>0$ are inessential positive constants and $q\geq 2N/(N-2)$.

The set-up of continuous $U(\Phi)$ with $U(0)=0$, along with conditions (R2)-(R3) (and $\displaystyle{U\in C^1\left(R^m\setminus\{0\}\right)}$ in (R4)(ii), (R4)(iii)), are guaranteed by our initial assumptions (A1)-(A4).

Condition (R1) requires the potential to be either stabilised -- that is, positive -- far away in field space, or, if it admits a runaway, the runaway slope is bounded by $2N/(N-2)$. In particular, for $N=4$, a potential of the form $\displaystyle{\lim_{|\Phi|\to\infty}U(\Phi)\sim-\lambda|\Phi|^4}$ formally fails (R1).

Conditions (A1)-(A4), (R1)-(R3), and (R4)(ii-iii) are satisfied, for example, 
by the polynomial potentials of Ref.~\cite{Dienes:2008qi}, as well as by many other supersymmetric potentials. 

A common exercise in the QFT literature is to study effective finite-order polynomial potentials, assumed to represent an expansion in the vicinity of some false vacuum. 
The quartic potentials studied in~\cite{Aravind:2014aza,Greene:2013ida,Dine:2015ioa} give recent examples. Allowing a quartic runaway, these potentials formally violate (R1). However, the discussion in these models (and many other examples) is limited in the first place to a finite region in field space, $|\Phi|<\Lambda$. To analyze ``little bounces" constrained to lie within $|\Phi|<\Lambda$, we are free to deform the potential such that $U(\Phi)>0$ for $\Phi>\Lambda$,  satisfying (A1)-(A4) and (R1)-(R4).

\section{Summary}\label{sec:conc}
We have considered scalar multi-field solutions of the Euclidean equations of motion (EOM). 
The reduced problem is defined as the problem of finding a field configuration vanishing at infinity that minimizes the kinetic energy $T$ at some fixed negative potential energy $V$. Ref.~\cite{Coleman:1977th} proved that, for $N>2$ Euclidean dimensions, if a solution of the reduced problem exists, then for appropriately chosen value of $V$ it is a minimum action solution of the EOM. It is the {\it bounce}~\cite{Coleman:1977py}, and dominates the decay of the false vacuum.
Ref.~\cite{Coleman:1977th} further showed that, for a single scalar field, the bounce is invariant under O($N$) rotations around its centre. To our knowledge, a proof of O($N$) symmetry of the bounce in the multi-field case has not been reported in the QFT literature. 

We made progress towards closing this gap and proved that if a solution of the reduced problem exists, then it and the minimum action solution of the multi-field equations derived from it are indeed O($N$) symmetric. We reviewed complimentary results from the math literature~\cite{Lopes1996378,2008arXiv0806.0299B}. The task of finding a proof of existence for a solution of the reduced problem was addressed in~\cite{brezis1984}, leading to a positive answer -- for example -- for finite-order polynomial potentials that are stabilised at $|\Phi|\to\infty$.

Interesting related questions include: (i) we have considered only canonical kinetic terms. What happens to the answer when  more general kinetic terms are allowed? (ii) what happens to the answer when gravity is included? (iii) while the minimum-action bounce is O($N$)-symmetric, an actual bubble nucleating in some cosmological set-up would not be, due to quantum fluctuations. How does one quantify the deviation from sphericity?

\section*{Acknowledgements}
We thank Yakar Kannai, Zohar Komargodski, Yossi Nir, Diego Redigolo, Steve Schochet, Adam Schwimmer, Amit Sever, Itai Shafrir, and Ofer Zeitouni for discussions.
The work of MT is supported by the JSPS Research Fellowship for Young Scientists.
KB is incumbent of the Dewey David Stone and Harry Levine career development chair, and is supported by grant 1507/16 from the Israel Science Foundation and by grant 1937/12 from the I-CORE program of the Planning and Budgeting Committee and the Israel Science Foundation.

\bibliography{ref}
\bibliographystyle{utphys}

\end{document}